# AN RFID-BASED CLINICAL INFORMATION SYSTEM FOR IDENTIFICATION AND MONITORING OF PATIENTS

Cristina Turcu[1], Tudor Cerlincă[2], Marius Cerlincă[3],
Remus Prodan[4], Cornel Turcu[5], and Felicia Gîză[6]

[1-6] Stefan cel Mare University of Suceava;
e-mail: {cristina, tudor_c, mariusc, prodan, cturcu, felicia}@eed.usv.ro

**Abstract** — Managing health-care records and information is an imperative necessity. Most patient health records are stored in separate systems and there are still huge paper trails of records that health-care providers must keep to comply with different regulations. This paper proposes an RFID-based system, named SIMOPAC, that integrate RFID technology in health care in order to make patient emergency care as efficient and risk-free as possible, by providing doctors with as much information about a patient as quickly as possible. Every hospital could use SIMOPAC with their existing system in order to promote patient safety and optimize hospital workflow. We will concentrate on the RFID technology and how it could be used in emergency care. We describe a general purpose architecture and data model that is designed for collecting ambulatory data from various existing devices and systems, as well as for storing and presenting clinically significant information to the emergency care physician.

## 1. INTRODUCTION

Electronic medical records (EMRs) may easily alleviate the distress of most doctors and nurses working in today's care system. Besides improving the degree of data availability among physicians and patients, they certainly increase the traceability of numerous medical details so deeply buried in traditional records.

Most hospitals have improved patient care by reducing wait times in the emergency ward when they decided to replace their paper-based process for emergency ward admission with a solution based on informatics systems. With these solutions in place, hospitals save minutes each time they admit a patient because doctors and nurses no longer fill out forms manually and improve healthcare outcomes. Although doctors have been slow to adopt electronic health records, their potential benefits cannot be disregarded. Thus, it has been estimated that 15 to 18 per cent of US physicians already use electronic health records [1].

"Instant access to patient information is key to lifesaving care, especially in the emergency room and intensive-care unit, where delays may mean the difference between life and death," Dr. Mark Smith said [3]. Currently, Emergency Medical Service (EMS) providers rely completely on personal and medical history information provided by patients or family members. It is common knowledge that stress, physical and mental discomfort prevent most patients and family members to impart vital medical information.

Problems may also arise if there are no family members around or if the patient is unconscious, incoherent or unable to talk or communicate (e.g. language difficulties). The next step beyond the EMR is to connect and provide medical information to primary care physicians, medicine and surgery specialists, anesthesiologists, nurse practitioners, assisted-living staff, patients themselves, patient's family and so on.

However, each hospital may use a different system, store data in many ways and even decide upon its own data format. Furthermore, file system access and data retrieval are often governed by inconsistent parameters seriously affecting the availability of medical information. Hence the inefficient communication among physicians.

Microsoft's Feied, a pioneer in medical training computer programs and medical intelligence software, said physician collaboration is the critical element for improving health care. He offered an impassioned testimonial. An emergency room physician who estimates he treated 80,000 patients "with my own hands," Feied said the thing that stuck out as he looked back on his career was how many times he was put in a position of "guessing over and over," "flying solo," in an information vacuum. In situations where people "die right in front of you," he said he often felt he was "one data element away" from stopping a patient from dying [1].

The market for bringing healthcare data from disparate sources into one view is growing by leaps, according to a new study from KLAS, a healthcare research firm based in Orem, Utah [4].

For example, through Microsoft HealthVault and Google Health, Microsoft and Google have a common goal of managing vast quantities of personal health information to benefit end users. Thus, these systems encourage and support healthcare patients/consumers to control and account for their own and family health records.

According to [2], data integration – the automated aggregation and consolidation information from a variety of disparate systems and sources – across sites of care (inpatient, ambulatory, home), across domains (clinical, business, operational), and across technologies (text, video, images) – is the Holy Grail of healthcare information technology.
But, it's necessary to find a way to get the vital medical data into the hands of those who can use it to save lives in emergency medical services, even when there is no connection to the Internet or the server is down.

## 2. RFID TECHNOLOGY

RFID technology is classified as a wireless Automatic Identification and Data Capture (AIDC) technology that can be applied for the identification and tracking of entities. An RFID device called RFID tag or transponders can be used as a means of identification. This tag contains an integrated circuit for storing information (including serial number and desired data), modulating and demodulating a (RF) signal, and other specialized facilities. The RFID tag transmits data in response to an interrogation received from a read-write device called RFID reader or interrogators. The tags and readers are designed with a specific operating frequency. Given the wireless communication between the RFID chip and the RFID reader, all data may be read from a distance. Tags fall into three categories: active (battery-powered), passive (the reader signal is used for activation) or semi-passive (battery-assisted, activated by a signal from the reader). Generally speaking, tag memory size can vary from 1 bit to 32 kbits and more. In certain tag types, the information on the tag is reprogrammable.

The general architecture of an RFID-based system is presented in Figure 1.

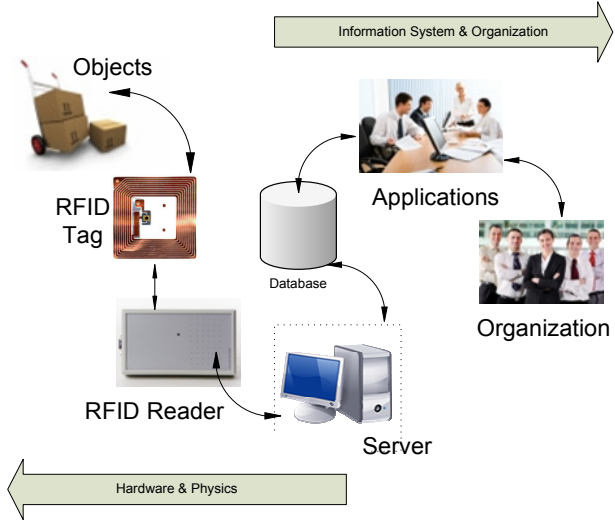

**Figure 1. The general RFID-based system architecture.**

The RFID technology has been used for inventory tracking, animal and person identification, supply chain management, toll collection and electronic payments, and vehicle identification for gate entry. For example, one important RFID based application is the e-passport. Although the photo page (including photo and physical characteristics of the person) was kept there are new additional features like the possibility to keep track of a person's travel history.

In today's healthcare applications, RFID technology is usually employed to:

- improve patient monitoring and safety;
- increase asset utilization with real-time tracking;
- reduce medical errors by tracking medical devices;
- enhance the efficiency of supply-chain.

For example the pharmaceutical industry has adopted this technology, particularly for the management of returns, contraindications, and product diversion and counterfeiting. For example, the US Food and Drug Administration (FDA) has launched a program to fight drug counterfeiting based on the use of RFID technology in packaging [5].

## 3. SIMOPAC PRESENTATION

We propose to develop a comprehensive clinical information system called SIMOPAC, based on RFID technology, for the rapid integration, organization, display and mining of data in real time from different EMR systems across regions. Thus, our proposed system could help link the different elements into a powerful system interconnecting hospitals, patients, health insurance companies and holders, public health officials and other health-related bodies.

Our system will use a passive RFID tag, named *Personal health Information Card* (CIP – in Romanian), to store the vital health information, such as medications, allergies and sensitivities. We consider the health information that is crucial in helping determine the way emergency physician should treat the patient. The system will allow physicians to retrieve valuable medical information just in time for a quick and accurate diagnosis.

Furthermore the CIP card will also store some data that will serve as a reference to patient electronic health care record from an informatics healthcare system, located elsewhere on the Internet, like URI (uniform resource identifier) of EMR server and patient identification information. The information that will be stored on the tag will allow the retrieval of medical information from the corresponding centralized database; this information is collected from many individual sources and to be effectively analyzed it must be characterized by consistency. Also, in order to protect the rights of both patients and providers, security and privacy mechanisms will be implemented.

The proposed RFID-based system could be used to ensure the positive patient identification (PPI) in a hospital. Our intention is to extend the procedure of patient identification beyond hospital and country boundaries. Thus, our RFID-based software system will be an open-loop RFID application functioning across global hospital boundaries. It will require the adoption of common standards (e.g. HL7) or new standards to support an international health information communication, especially for emergency healthcare.

The system will be developed for a greater interoperability of electronic medical records and personal health records. It is an inspired technical solution offered to healthcare professionals in order to preserve and transfer complex healthcare and patient information that requires no replacement or change of the informatics system employed.

## 3.1. SIMOPAC Architecture

The SIMOPAC architecture is shown in Figure 2.
Since the system architecture will allow users to capture, integrate and display an enormous amount of patient-related data from a variety of sources, medical decisions and outcomes will be considerably improved. The CIP will allow the identification of patients, and the RFID card will provide access to an ambulatory EMR, namely a data repository devised as a subset of a longitudinal health record.

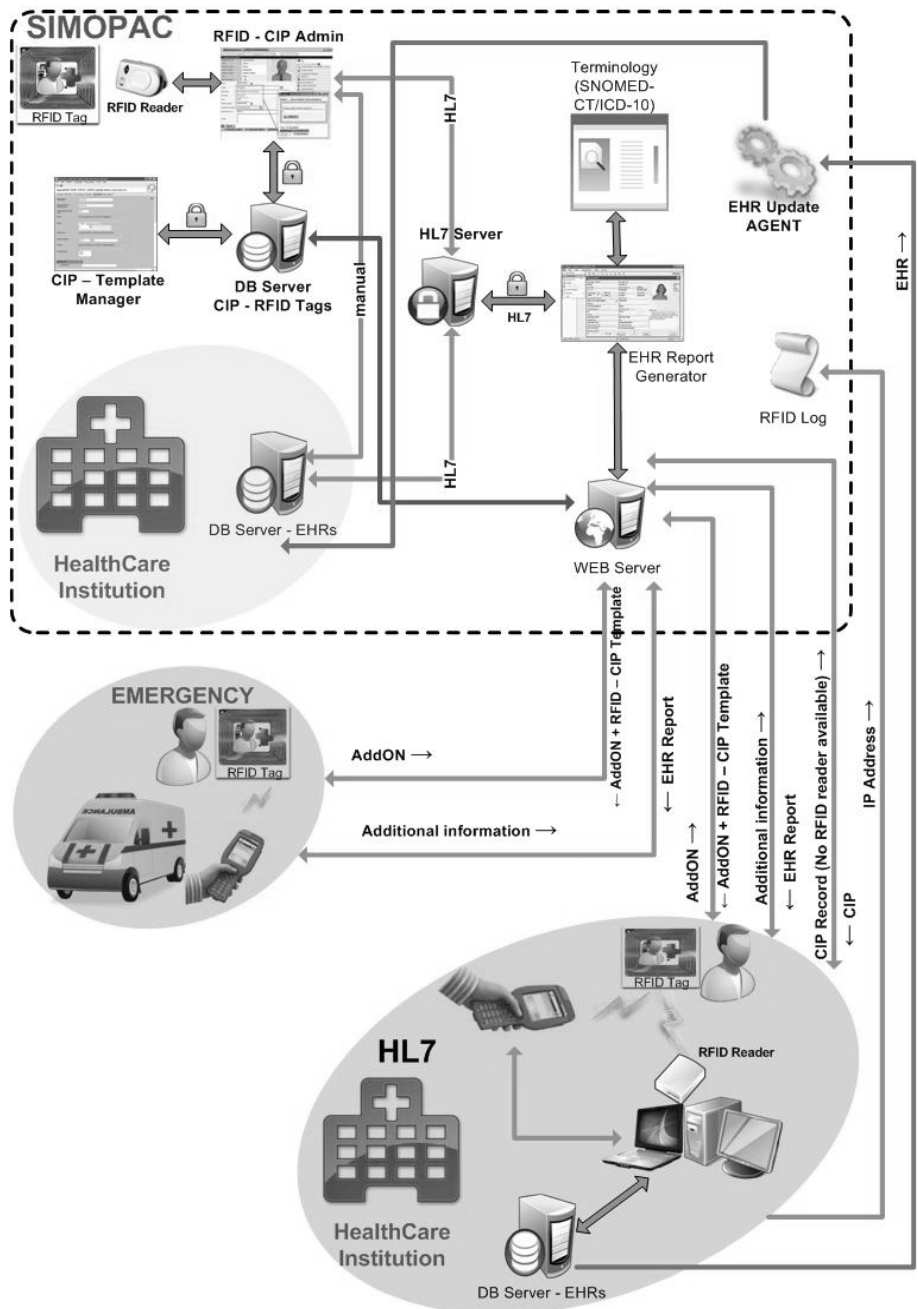

**Figure 2. The SIMOPAC architecture.**

Furthermore, the CIP could be used to allow physicians to connect to the SIMOPAC server. To link patient identifiers to patient information the SN@URI approach has been proposed, where SN represents the CIP serial number and URI stands for the uniform resource identifier.

Through SIMOPAC physicians could define a template using a wide variety of data types in order to provide a composite portrait of their patients' healthcare history. The role of the template manager module is to support the whole template management. All patients' values corresponding to the template fields will be stored on the CIP.

Thus, in order to provide an up to date patient's healthcare history, SIMOPAC will tie together many unrelated EHR systems that are using a wide variety of data types.

The RFID medical record has been devised in such a way as to provide a synopsis of patient healthcare information and, more importantly, to store only those values corresponding to significant data defined by the template. It is meant to enable a rapid assessment of patients' overall health and recent visits to healthcare providers.

For every system that produces patient data, SIMOPAC will create an agent collecting data according with HL7 standard. The multi-agent system will be able to:
- gather healthcare information electronically across healthcare units;
- bring real-time data from a wide variety of clinics and hospital departments into a concise collection of information that can be shared.

In this way a clinician may visualize the medical history of any listed patient. Nevertheless, the data from the original sources is not co-mingled.

Thus, through SIMOPAC when a physician refers a patient to a hospital, he is also able to provide an electronic copy of some relevant part of his/her patient's medical record extracted from the electronic medical record (EMR). In addition, when the patient is discharged from a hospital, a discharge summary could be transferred from the facility's hospital information system (HIS)/EMR into the doctor's system through the CIP and the agent system. That would allow the electronic loop to be closed.

Furthermore, SIMOPAC will provide a translator, so no manner which places the physician is working he could understand the data presented. Thus, SIMOPAC will enable the querying of different sections of a single user's medical record or across multiple user records, even in different EHR systems across the globe.

Terminology module will provide service for validation, browsing and translation of HL7 specific codes. Also, this module will provide terminology services for external coding systems such as ICD-9, ICD-10, national drug code lists.

The URI on the CIP enables a rapid assessment of the patient recent visits to healthcare providers. Using the URI on the CIP SIMOPAC allows physician to obtain historical patient data instantly. It communicates with numerous different hospital computer systems and automatically pulls up comprehensive medical information about individual patients identified through the CIP.

In this new collaboration system of health information providers, information privacy and control will become an important issue. Thus, privacy and interoperability issues are potential deal breakers as the medical practice goes electronic. SIMOPAC access will be secure, password protected. Furthermore, there will be many checks and balances in the system to ensure that patient confidentiality is safeguarded and that only people eligible are granted system access. Our system will implement the highest standard of web security involving the use of digital certificates and end-to-end public key encryption.

SIMOPAC will be able to scale across a range of devices, including Pocket PCs, and to handle a large number of work stations and a large volume of medical data. Thus PDAs will be used by emergency physicians at the emergency scene to input data. An RFID reader device attached to a handheld will allow the emergency physician to identify a patient and read the stored vital information about the patient. In addition, the handheld device will be equipped with 802.11b wireless connectivity, to allow the medical staff mobility and network connectivity to the hospital's information systems.

The system will provides an easy-to-use interface through which medical users can see any patient's medical history with just a few clicks.

We will use Java and postgreSQL database software to create a Web-accessible interface so that physicians may see their patients' medical records via an Internet connection from anywhere in the world.

SIMOPAC will gather data from all sources of information and merge them together.

### *3.2. BENEFITS*

The great advantage of SIMOPAC is represented by the availability of vital medical information through patient cards and the usage of health information standards (such as HL7) to store, transfer and retrieve users' health records. All technical solutions have subordinated to two major goals:
- Healthcare providers can give patients and their family members personal cards with their critical health information;
- Healthcare software and device developers can develop personalized services for individual users by employing the patient's health information in the system.

The development and deployment of SIMOPAC offers numerous benefits. Thus, it
- ensures the electronic access to clinical information among disparate health care information systems;
- facilitates the access to and the retrieval of clinical data with the view to providing safer, more timely, efficient, effective, equitable and patient-centered care;
- can reduce redundant procedures, tests, etc.
- sees all data easily, using a simple and user-friendly interface;
- eliminates transcription;
- helps public health authorities to perform rapid analyses and produce health reports on any population segment.

## 4. CONCLUSIONS

Most hospitals store patient medical data that cannot easily be shared with other systems because of disparate data types. We propose an RFID-based integrated system that will aggregate health-related data across more hospitals according to recognizable standards; it will make it available for emergency departments in hospitals and public health officials. This solution could help save lives by giving hospitals and paramedics up to date information, on scene, when responding to a medical emergency situation. Thus, in emergency medical situations, SIMOPAC offers EMS providers life-saving medical information via the CIP containing up-to-date medical information of vital importance to those entitled to make fast and accurate patient care decisions.

The SIMOPAC complexity is further amplified by the fact that most individual electronic health record systems are packaged products supplied by a variety of independent software vendors and run on different platforms.

## ACKNOWLEDGMENT

The research results and technical solutions presented in this paper have received the support of Grant named "SIMOPAC – Integrated System for the Identification and Monitoring of Patient" no. 11-011/2007 within the framework of the Romanian Ministry of Education and Research "PNCDI II, Partnerships".